\documentclass[%
	       preprint2
  ]{aastex}

\usepackage{amsmath} 
\usepackage[utf8]{inputenc}
\usepackage{epsfig}
\usepackage{url}

\usepackage{color}
\usepackage[colorlinks=true]{hyperref}

\defcitealias{korn2007}{Paper~I}
\defcitealias{lind2008}{Paper~II}

\newcommand\red[1]{{\color{red} #1}}
\newcommand\draftbf[1]{\textbf{#1}} 
\renewcommand\draftbf[1]{#1}
\newcommand\priv{private comm.}
\newcommand\coloredition{[{\it See the electronic edition of the Journal for a color version of this figure.}]}
\renewcommand\coloredition{}

\newcommand{\tme}{turbulent-mixing efficiency} 
\newcommand{\tmes}{turbulent-mixing efficiencies} 
\newcommand{\pretme}{\tme} 

\newcommand\ts[1]{_{\rm #1}} 
\newcommand\figref[1]{\red{\autoref{#1}}} 
\newcommand\tblref[1]{\red{\autoref{#1}}} 
\newcommand\secref[1]{\S~\ref{#1}}

\newcommand\scaleshifted{{\rm '100s'}}
\newcommand\scaleone{{\rm '100e'}}
\newcommand\scaletwo{{\rm '200e'}}

\newcommand\Halpha{{\rm H$\alpha$}}

\newcommand\thecluster{{\rm NGC 6397}}
\newcommand\hdref{{\rm HD 84937}}
\newcommand\hdsub{{\rm HD 140283}}
\newcommand\fieldpoor{{\rm G64-12}}
\newcommand\wmap{{\rm WMAP-7}}
\newcommand\elem[1]{{\rm #1}}
\newcommand\abundance{\ensuremath{\log \varepsilon}}
\newcommand{\Teff}[1][]{\ensuremath{T_{\rm eff#1}}} 
\newcommand{\logg}[1][]{\ensuremath{\log g_{\rm #1}}} 
\newcommand\microturb{\ensuremath{\xi}}
\newcommand\meanthreed{\ensuremath{\langle {\rm 3D} \rangle}}
\newcommand\TOP{{\rm TOP}}
\newcommand\SGB{{\rm SGB}}
\newcommand\bRGB{{\rm bRGB}}
\newcommand\RGB{{\rm RGB}}
\newcommand\NLTE{{\rm NLTE}}
\newcommand\FeH{\ensuremath{[\elem{Fe} / \elem H]}}
\newcommand\SI[2]{#1\,\text{#2}} 

\newcommand\stagger{the {\sc Stagger Code}}

\begin{document}

\title{Atomic Diffusion and Mixing in Old Stars. III. 
Analysis of NGC~6397 stars under new constraints}
\author{T. Nordlander and A. J. Korn} 
\affil{Department of Physics and Astronomy, Division of Astronomy and Space Physics, Uppsala University, Box 516, 75120, Uppsala, Sweden}
\author{O. Richard}
\affil{LUPM, Universit\'e Montpellier 2, CNRS, place Eug\`ene Bataillon, 34095 Montpellier, France}
\and
\author{K. Lind}
\affil{Max-Planck-Institut f\"ur Astrophysik, Karl-Schwarzschild-Strasse 1, 857 41, Garching bei M\"unchen, Germany}

\shorttitle{Atomic Diffusion and Mixing in Old Stars III.}
\shortauthors{Nordlander et al.}

\begin{abstract} 
 We have previously reported on chemical abundance trends with evolutionary state in the globular cluster NGC~6397 discovered in analyses of spectra taken with FLAMES at the VLT.
 Here, we reinvestigate the FLAMES-UVES sample of 18 stars, ranging from just above the turnoff point (TOP) to the red giant branch below the bump.
 Inspired by new calibrations of the infrared flux method, we adopt a set of hotter temperature scales. 
 Chemical abundances are determined for six elements (Li, Mg, Ca, Ti, Cr, and Fe).
 Signatures of cluster-internal pollution are identified and corrected for in the analysis of Mg.

On the modified temperature scales, evolutionary trends in the abundances of Mg and Fe are found to be significant at the $2 \sigma$ and $3 \sigma$ levels, respectively.
 The detailed evolution of abundances for all six elements agrees with theoretical isochrones, calculated with effects of atomic diffusion and a weak to moderately strong efficiency of turbulent mixing. 
 The age of these models is compatible with the external determination from the white dwarf cooling sequence.
 We find that the abundance analysis cannot be reconciled with the strong turbulent-mixing efficiency 
 inferred elsewhere for halo field stars.
 A weak mixing efficiency 
 reproduces observations best, indicating a diffusion-corrected primordial lithium abundance of $\log \varepsilon ({\rm Li}) = 2.57 \pm 0.10$. At $1.2 \sigma$, this value agrees well with WMAP-calibrated Big-Bang nucleosynthesis predictions.
\end{abstract}


\section{Introduction}

NGC~6397 is one of the best-studied metal-poor ($\FeH \,\approx\, -2.1$) globular clusters of the Milky Way. Prompted by recently determined external constraints detailed below, we continue and update our investigation of stars in this cluster. In a series of articles -- \citet{korn2006a}, \citet[hereafter \citetalias{korn2007}]{korn2007} and \citet[hereafter \citetalias{lind2008}]{lind2008} -- homogeneous, internally consistent analyses of stars at various evolutionary stages were performed on spectra taken with the multi-object spectrograph FLAMES at the VLT. 
In \citetalias{korn2007}, the evolution of surface elemental abundances was traced from high-resolution FLAMES-UVES observations of 18 stars just past the turnoff-point (TOP), on the subgiant branch (SGB), at the base of the red giant branch (bRGB), and on the red giant branch below the bump (RGB). 
The elemental abundances of Li, Mg, Ca, Ti and Fe were found to display systematic evolutionary trends incompatible with standard stellar-evolution models, significant  at the $3 \sigma$ level in the cases of Mg and Fe. 
Comparing with stellar-evolution models including the effects of atomic diffusion as calculated by \citet{richard2002,richard2005}, the trends were found compatible with those of theoretical isochrones for an age of $\le\,$\SI{13.5}{Gyr}, \draftbf{with turbulent mixing of moderate efficiency}.
These results were supported by the independent analysis in \citetalias{lind2008}, where abundances were derived for Fe, Ti, Ca and Mg from medium-high resolution GIRAFFE observations of 116 stars along the subgiant branch in the same cluster.

\draftbf{A number of recent studies have set new constraints on the 
temperature scale of metal-poor stars, 
cluster-internal pollution, 
the age of NGC~6397 
and the 
efficiency of the turbulent mixing which moderates the effects of atomic diffusion. 
It is the aim of this article to reevaluate the abundance trends and their comparison to model predictions under these new constraints.}

\draftbf{On the observational side}, the improved IRFM calibration of \citet{casagrande2010} for dwarfs and subgiants produces $\sim\,\SI{100}{K}$ higher effective temperatures than the \Halpha\ analyses of \citetalias{korn2007}.
Similar alterations to the temperature scale were proposed by \citet{bonifacio2007}, as they would significantly reduce the strength of observed trends thus making the results of \citetalias{korn2007} agree better with the analyses of \citet{gratton2001} who found no abundance differences between TOP and bRGB stars.

Na-O anticorrelations as inferred by \citet{carretta2009} from a very large set of red giants in 15 globular clusters, indicate a largely bimodal distribution of primordial (first-generation) and polluted (second-generation) stars.
In \thecluster, the primordial component in a sample of 16 RGBs with reliable oxygen abundances is found to encompass roughly 25\,\% of all stars, similar to other globular clusters of the same study. 
\draftbf{For the same cluster, \citet{lind2009b,lind2011} find a similar distribution of sodium abundances, tied to significant anticorrelations with e.g. magnesium.}

The age of \thecluster\ has been determined in analyses of the white dwarf cooling sequence (hereafter WDCS) by \citet{hansen2007} and \citet{kowalski2007}. We adopt the latter's determination as our WDCS age prior, $12.0 \pm \SI{0.5}{Gyr}$ ($2 \sigma$), to check whether the results of \citetalias{korn2007} are modified by this boundary condition \draftbf{on the theoretical modeling}.

The study of \citet{gonzalez2009} compares 3D-NLTE lithium abundances of main-sequence and subgiant stars at similar temperature in \thecluster. \draftbf{They find a behavior on the upper subgiant branch in rough qualitative agreement with atomic-diffusion model predictions employing a high \pretme.
\citet{melendez2010} find good correlation with the mass-dependence of lithium depletion in field halo stars for diffusion models using the same (high) \pretme.} Their deduced primordial lithium abundance, $\abundance(\elem{Li}) = 2.64$, is in excellent agreement with predictions of the Big-Bang nucleosynthesis (BBN) lithium abundance using current \wmap\ and nuclear reaction data and uncertainties,
$\abundance(\elem{Li}) = 2.71 \pm 0.06$ \citep{cyburt2010}. 
\draftbf{We evaluate whether models adopting such high \pretme\ predict significant abundance variations in elements other than lithium, compatible with observations.}

In the present work, we revisit the stellar sample and spectra of \citetalias{korn2007}.
We introduce three new temperature scales in \secref{sec:tscales}.
\draftbf{We present our redetermined elemental abundances on the new temperature scales} in \secref{sec:results}, and our efforts to correct for effects of pollution in \secref{sec:pollution}.
We compare these results to a grid of theoretical isochrones with effects of atomic diffusion at various ages in \secref{sec:diffusion}. 
We summarize and consider the implications for the cosmological lithium problem in \secref{sec:discussion}.

\section{Methods} \label{sec:methods}
\begin{deluxetable}{lll}
\tablecaption{Line list used in updated abundance determinations\label{tbl:lines}}
\tablewidth{0pt}
\tablehead{
  \colhead{Species} & \colhead{Line(s) [\AA]} & \colhead{Treatment}
}
\startdata
\ion{Li}1 & $\lambda$6707 & NLTE \\
\ion{Na}1 & $\lambda \lambda$5688, 8183\tablenotemark{a}, 8194\tablenotemark{a} & NLTE \\
\ion{Mg}1 & $\lambda$5528 & NLTE \\
\ion{Ca}1 & $\lambda \lambda$6122, 6162, 6439 & NLTE \\
\ion{Ti}2 & $\lambda \lambda$5188, 5226 & LTE \\
\ion{Cr}1 & $\lambda$5206 & NLTE \\
\ion{Fe}2 & $\lambda \lambda$4923.9, 5018.4, 5169.0, 5316.6, 5234.6 & LTE
\enddata
\tablenotetext{a}{IR lines from supplemental spectra \citep{gratton2001,lind2009b}.}
\end{deluxetable}

Photometric and spectroscopic analyses are performed as detailed in \citetalias{korn2007}.
We start from the spectroscopically determined, internally consistent stellar parameters determined there:
MAFAGS-ODF 1D LTE atmospheres \citep{grupp2004} were used to iteratively derive \Teff, \logg, \microturb, and \FeH\ for each star. \Teff\ was determined using \Halpha\ wing fitting employing \citet{ali1966} broadening theory, \logg\ from the iron ionization equilibrium in NLTE \citep{korn2003}, 
and \microturb\ from the \elem{Fe} line-strength trend.

We deviate from this internally consistent analysis by instead adopting absolute \logg\ values derived from photometric $V$ magnitudes, using the same parameters for reddening and distance modulus as in \citetalias{korn2007}. 
We employ temperatures from the new temperature scales defined in \secref{sec:tscales}. 
Bolometric corrections are adopted from \citet{worthey2011}.
Stellar masses are taken from a 12.5~Gyr isochrone with atomic diffusion. 
\draftbf{Models without atomic diffusion would differ only by an absolute offset in \logg.}
On the TOP and SGB, the expected $\sim \SI{0.25}{dex}$ underabundance in helium due to diffusion alters the mean molecular weight, corresponding to a decrease in spectroscopic \logg\ values \citepalias[see][]{korn2007}.
We account for this by decreasing \logg\ values by \SI{0.05}{dex} in our spectroscopic analysis of TOP and SGB stars. This assumption affects only the abundances of Ti and Fe (derived from ionic lines), \draftbf{decreasing both by a mere \SI{0.01}{dex}}.


Chemical abundances are deduced from line-by-line solar-differential profile fits, thus cancelling the dependence on adopted $\log gf$ values. 
Group average chemical abundances are not averages of individual stellar abundances -- rather, they are derived from mean stellar parameters applied to group averaged spectra. \draftbf{This approach allows us to work on spectral lines as weak as 20 m\AA, as is the case for the ionic lines of titanium in the TOP stars.}
The line list employed in this updated analysis is given in \tblref{tbl:lines}. 
Abundance uncertainties on the differential scale are assumed to be \SI{0.05}{dex}, save for the cases of \elem{Mg} (see \secref{sec:pollution}) and \elem{Fe} \citepalias[see][Table~3]{korn2007}.

\subsection{NLTE and 3D modeling} \label{sec:nlte}
\draftbf{We make an effort to employ the most realistic line-formation theory available. 
NLTE corrections are applied for} \ion{Mg}1 \citep{gehren2004}, \ion{Ca}1 \citep{mashonkina2007} and \ion{Cr}1 \citep{bergemann2010}. Additionally, we apply NLTE corrections to LTE-derived abundances for \ion{Na}1 \citep{lind2011b} and \ion{Li}1 \citep{lind2009a}. The majority species \ion{Fe}2 and \ion{Ti}2 are modeled in LTE, as they have been shown to exhibit only minor NLTE effects \citep{mashonkina2011,bergemann2011}.

We derive iron abundances from \ion{Fe}2 lines only, as the modeled strengths of \ion{Fe}1 lines depend critically on NLTE and its modeling assumptions (specifically the inelastic \ion H 1 collision factor $S \ts H$).
As a diagnostic of this effect, we have also derived \logg\ values from the iron ionization equilibrium on our new temperature scales. For this test, we employ NLTE corrections in the line formation producedure following \citet{korn2003}, with rather efficient hydrogen collisions ($S_{\rm H} = 3$). 
\draftbf{These spectroscopically derived \logg\ values overall agree reasonably with those derived directly from photometry and distance.} 
A significant difference is found on our hottest temperature scale, \scaletwo, where \logg\ values for the TOP stars are higher by \SI{0.2}{dex}, placing them at $\logg = 4.2$, which is just below the turnoff point in our theoretical CMD (compare to \figref{fig:kiel}).

3D effects on abundance trends in LTE were investigated in \citetalias{korn2007}. Qualitative results implied stronger trends for \ion{Ti}2 and \ion{Fe}2, but a weaker trend for \ion{Mg}1.

For chromium, we find similar NLTE corrections for the high-excitation triplet line \ion{Cr}1 $\lambda 5206$ in all groups, ranging from \SI{0.31}{dex} at the RGB to \SI{0.37}{dex} at the SGB. 
\citet{bonifacio2009} predict very large differential 3D effects at low metallicities. At $\FeH = -3.0$, these corrections comparing giants to dwarfs are on the order of $\delta \Delta \abundance (\elem{Cr}) =$ \SI{0.4---0.5}{dex} for high excitation lines.
Preliminary \meanthreed\ NLTE modeling implies individual corrections of similar magnitude \citetext{M. Bergemann, \priv}, rendering comparisons between dwarfs and giants uncertain.
Thus, we expect that our results may change significantly under more sophisticated modeling.

For lithium, we need not only consider differential modeling effects, but must also determine absolute abundances to  compare with the predicted primordial abundance. 
The 3D NLTE treatment by \citet{sbordone2010}, using an 8-level model atom, results in corrections $\Delta ({\rm 3D} - {\rm 1D}) = \SI{0.03}{dex}$ for our TOP and SGB stars. In 1D, these NLTE corrections ($\Delta \NLTE$) are very near zero.

We have performed independent analyses following \citet{lind2009a} on metal-poor stars similar to our sample: for the metal-poor dwarfs \hdref\ and \fieldpoor, and the subdwarf \hdsub\ representing well studied Spite plateau stars similar to our TOP and SGB stars, as well as an additional model representing a star evolved somewhat beyond our sample (at $\Teff = \SI{5000}K$, $\logg = 2.0$, $\FeH = -2.0$). In this analysis, 1D models were constructed from MARCS atmospheres \citep{gustafsson2008}.
Average 3D (\meanthreed) models were constructed from sequences of snapshots from radiation-hydrodynamic convection simulations using \stagger\ \citep{stagger}, with averaging over the 3D cubes on surfaces of equal $\tau_{500}$ \citetext{R. Collet, \priv}. 
Results imply \meanthreed\ NLTE lithium abundances very similar to the corresponding 1D NLTE case. On the Spite plateau, we find $\Delta \NLTE = \SI{-0.05}{dex}$ in 1D, with identical or slightly higher abundances in \meanthreed. 
For evolved stars, the correction instead becomes positive.
For the evolved model star investigated here, the NLTE abundance is lower in \meanthreed\ than in 1D. The NLTE effects in this star should be greater than in our bRGB and RGB stars, as it has significantly lower surface gravity. 
These results imply that hydrodynamical effects will not lead to significantly different results for our evolved stars, and that 1D LTE analyses indeed underestimate their lithium abundances.

\draftbf{Similarly small $\Delta ({\rm 3D}-{\rm 1D})$ corrections result from these preliminary 3D and \meanthreed\ NLTE lithium abundance analyses. For this reason, we feel safe in applying 1D NLTE corrections, but keep in mind the larger systematic uncertainty in the abundances of evolved stars.}

\subsection{Scenarios} \label{sec:tscales}

\begin{deluxetable}{lc cccc ccc cc}
 \tabletypesize{\small}
\tablecaption{Photometric effective temperatures\label{tbl:photometry}}
\tablewidth{0pt}
\tablehead{
  \colhead{Group} & \colhead{Spectroscopic} &
  \multicolumn4c{IRFM 1996/1999\tablenotemark{a}} &
  \multicolumn3c{IRFM 2010\tablenotemark{b}} &
  \multicolumn2c{MARCS\tablenotemark{c}} \\
\cline{3-6} \cline{7-9} \cline{10-11} 
\\[-2ex] 
  & \colhead{\Halpha} &  $b-y$ & $v-y$ & $B-V$ & $V-I$ &    $b-y$ & $B-V$ & $V-I$ &          $b-y$ & $v-y$
}
\startdata
  TOP  & 6254 & 6229 & 6213 & 6160 & 6133 & 6374 & 6373 & 6327 & 6288 & 6281
\\SGB  & 5805 & 5797 & 5824 & 5827 & 5688 & 5951 & 6004 & 5804 & 5839 & 5853
\\bRGB & 5456 & 5396 & 5408 & 5545 & 5317 & \dots & \dots & \dots & 5541 & 5535
\\RGB  & 5130 & 5121 & 5132 & 5254 & 5063 & \dots & \dots & \dots & 5225 & 5226
\\[.8ex] \hline \\[-1ex] 
$\Delta(\TOP-\RGB)$ & 
1124 & 1108      &  1082     &  \phn906      &   1070    & \dots & \dots & \dots & 1063 & 1055 \\
$\Delta(\TOP-\SGB)$ &
\phn449  &  \phn433  &  \phn390  &  \phn333  &  \phn446  & \phn 423 & \phn 369 & \phn 523 & \phn449  & \phn428
\enddata
\tablecomments{Results have been averaged for each group. Photometric calibrations are evaluated for $\FeH = -2.0$, on measurements listed in \citetalias[Table~9]{korn2007}.}
\tablenotetext{a}{Empirical calibrations for dwarfs \citep{alonso1999} and giants \citep{alonso1996}. We consider the SGB stars to belong to the dwarf calibration, and the bRGB stars to the giant one.}
\tablenotetext{b}{Empirical calibrations valid for dwarfs and subgiants \citep{casagrande2010}.}
\tablenotetext{c}{Synthetic photometry valid for dwarfs and giants \citep{onehag2009}.}
\end{deluxetable}

We compare the effective temperatures deduced from spectroscopic \Halpha\ analyses to those from three different photometric calibrations in \tblref{tbl:photometry}.
All photometric calibrations indicate a range $\Delta \Teff(\TOP-\RGB)$ similar to or somewhat more narrow than the spectroscopic analysis. Working on these temperature scales would generally strengthen the previously identified trends. 

The \citet{casagrande2010} calibrations for dwarfs and subgiants indicate higher effective temperatures for the TOP by $110 \pm \SI{30}K$, \draftbf{but are not applicable to giants}.
Assuming the empirical calibrations of \citet{alonso1996} hold on the RGB, this would allow an increase in $\Delta \Teff(\TOP-\RGB)$ by \SI{100}K. 
Note however that this is not the most likely interpretation of the hotter temperatures stemming from the \citet{casagrande2010} calibration.
Rather, \draftbf{the difference} is largely due to the calibration of the infrared bands, resulting in an absolute offset, rather than the inclusion of interferometric radii for dwarfs, which would not affect the giants and thus give a differential effect \citetext{L. Casagrande, \priv}.

On the SGB, the effective temperatures are similarly higher by $115 \pm \SI{100}K$, indicating no significant change in $\Delta \Teff(\TOP-\SGB)$. The errors are, however, fairly large, as the calibration sample is rather thin at this evolutionary stage \citep[see Fig.~13 of][]{casagrande2010}. 

The \Halpha\ analysis employs the broadening theory of \citet{ali1966}. Direct comparison to that of \citet{barklem2000} indicates only minor deviations, affecting $\Delta \Teff(\TOP-\RGB)$ by \SI{20}K, and increasing $\Delta \Teff(\TOP - \SGB)$ by some \SI{40}K.

Starting from the spectroscopic temperature scale, we introduce three hypothetical temperature scales. For each, photometric \logg\ values are determined as described in \secref{sec:methods}.
In the most probable scenario, we \draftbf{shift} all effective temperatures by \SI{100}K, to match the expected absolute offset introduced in the \citet{casagrande2010} calibration. We refer to this temperature scale as \scaleshifted.

Additionally, we create two expanded temperature scales.
The first retains $\Delta \Teff(\TOP-\SGB)$ in reasonable agreement with both photometry and spectroscopy, expanding the temperature scale linearly, by a fixed amount for each stellar group as determined by its mean temperature. 
We \draftbf{expand} $\Delta \Teff(\TOP-\RGB)$ by \SI{100}K, keeping temperatures fixed at the RGB, and refer to this temperature scale as \scaleone.
The increase in parameter $\Delta \Teff(\TOP-\SGB)$, \SI{40}K, is in agreement with the improved \Halpha\ broadening, and does not violate the photometric indicators. This scenario represents a realistic but conservative case, relative to \scaleshifted.
The temperature scale \scaletwo\ is created by the same method, with $\Delta \Teff(\TOP-\RGB)$ \draftbf{expanded} by \SI{200}K. This temperature scale, while poorly supported by our photometric and spectroscopic analyses, is meant to represent the case where atmospheric parameters are tuned to flatten the abundance trends. 


\begin{figure}[t]
\plotone{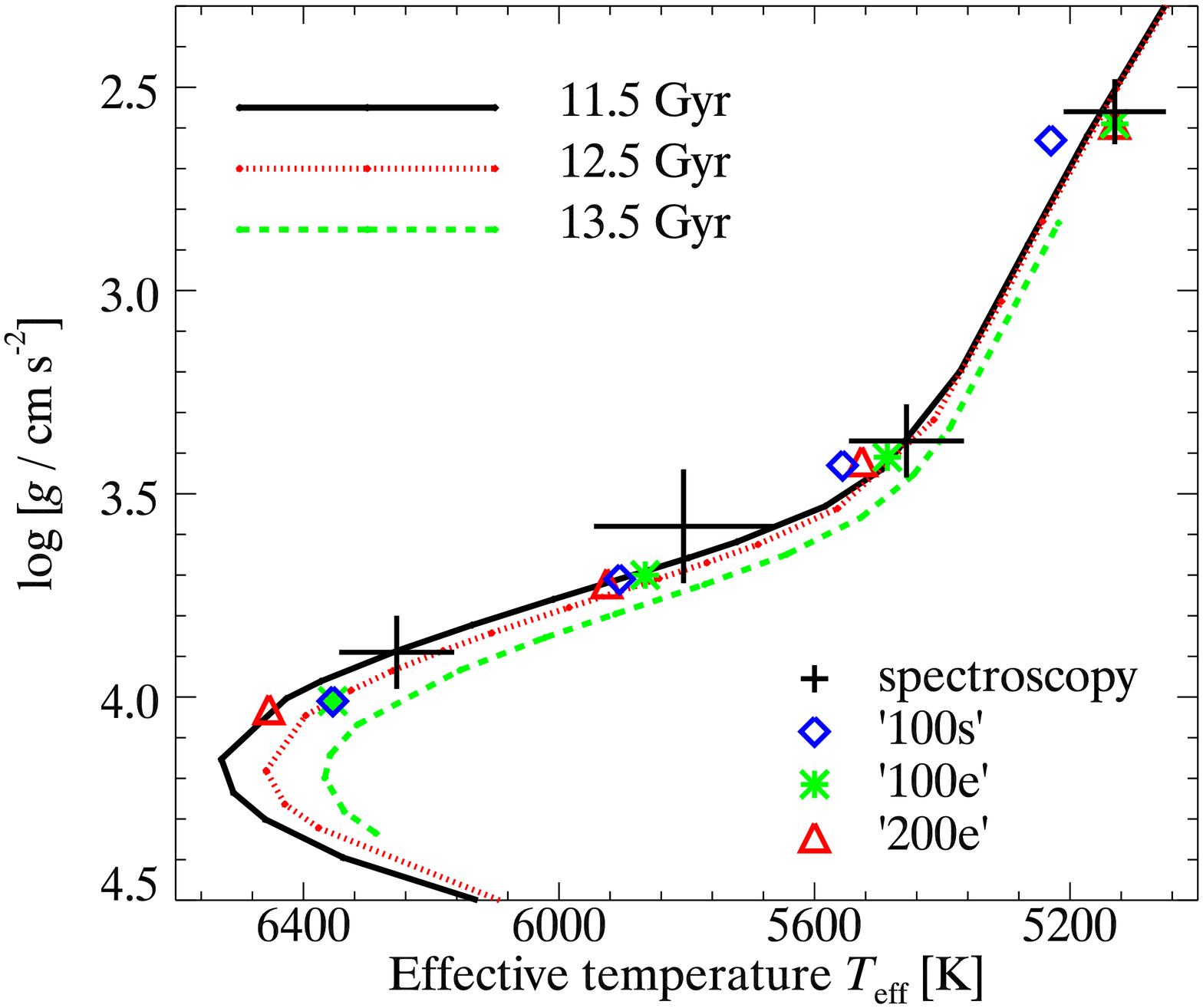}
\caption{Comparison of predicted stellar parameters 
for theoretical T6.09 isochrones and the stellar samples. 
Symbols represent the various theoretical temperature scales (see text).
Shown as crosses are results with error bars from the spectroscopic analysis of \citetalias{korn2007}. 
Note that the \SI{13.5}{Gyr} model was computed at \SI{0.1}{dex} higher metallicity, which produces a slight deviation on the RGB. The difference is negligible near the TOP.
\coloredition}
\label{fig:kiel}
\end{figure}

\draftbf{The stellar parameters on each temperature scale} are shown together with theoretical isochrones at three different assumed ages in \figref{fig:kiel}. 
The WDCS prefers an age of $\SI{12.0 \pm 0.5}{Gyr}$ \citep{hansen2007,kowalski2007}, \draftbf{which the models effectively bracket}.
This age constraint is well fulfilled by all three temperature scales.
Additionally, the spectroscopic values are compatible within their error bars.

An additional constraint comes from the observed cluster CMD \citepalias[see][Fig.~2]{korn2007}. 
We expect our TOP stars to lie some \SI{100---200}K cooler than the turnoff point \citetext{F. Grundahl, \priv}. 
This criterion disfavors the hot \scaletwo\ temperature scale as well as the spectroscopic determination -- indicating a cluster too young or old, respectively -- while \scaleshifted\ and \scaleone\ are both fully compatible.

\draftbf{For traditional stellar models computed without atomic diffusion, the lack of gravitational settling of helium results in a significantly older cluster.}
Then, only \scaletwo\ seems compatible with the WDCS age constraint, as well as the cluster CMD. From the cluster CMD constraint, \scaleshifted\ and \scaleone\ would both prefer \draftbf{ages greater than \SI{13.5}{Gyr}}.

\subsection{Stellar models with diffusion effects} \label{sec:diffmethod}
Following \citet{richard2002,richard2005}, a sparse grid of isochrones has been computed with effects of atomic diffusion. Three different ages and six different efficiencies of turbulent mixing are considered. At 11.5 and \SI{12.5}{Gyr}, a metallicity of $\FeH = -2.1$ was used. The \SI{13.5}{Gyr} models were reused from \citetalias{korn2007}, with $\FeH = -2.0$. This has minor effects on e.g. the stellar mass at a given evolutionary stage, but is negligible for the evolution of surface abundances. Turbulent mixing is prescribed ad hoc with an efficiency parametrized by a reference temperature $T_0$ determining the behavior in both strength and depth \citep[see section~3.2 of][for more details]{richard2005}. We use the shorthand T$x$ for models using $\log T_0 = x$, and refer to the magnitude of this parameter as the \tme. 
The T5.95 and T6.25 models respectively represent the lowest and highest \tmes\ that seem compatible with the flatness of the Spite plateau of lithium.

\section{Results} \label{sec:results}

Mean stellar parameters and group-average abundances are presented in \tblref{tbl:params}.

\begin{deluxetable}{lcccccccc}
\tablecaption{Mean stellar parameters and abundances at the different temperature scales\label{tbl:params}}
\tablewidth{0pt}
\tablecolumns{9}
\tablehead{
  \colhead{Group} & \colhead{\Teff} & 
  \colhead{\logg\tablenotemark{a}} & 
\colhead{$[\elem{Fe}/\elem H]$} & 
\colhead{$\abundance(\elem{Li})$} &
\colhead{$\abundance(\elem{Mg})$\tablenotemark{b}} &
\colhead{$\abundance(\elem{Ca})$} &
\colhead{$\abundance(\elem{Ti})$} &
\colhead{$\abundance(\elem{Cr})$}
\\ 
&
  \colhead{[K]} &
  \colhead{[cgs]} &
  \colhead{NLTE} &
  \colhead{NLTE} &
  \colhead{NLTE} &
  \colhead{NLTE} &
  \colhead{LTE} &
  \colhead{NLTE}
}
\startdata
  \cutinhead{The \SI{100}K shifted temperature scale \scaleshifted}
   TOP & 6354 &  4.01 & -2.23 &  2.26 &  5.80 &  4.59 &  3.05 &  3.41
\\ SGB & 5905 &  3.71 & -2.19 &  2.39 &  5.91 &  4.58 &  2.98 &  3.46
\\bRGB & 5556 &  3.38 & -2.13 &  1.46 &  5.92 &  4.63 &  3.05 &  3.49
\\ RGB & 5230 &  2.58 & -2.07 &  1.12 &  5.98 &  4.70 &  3.10 &  3.47
\\\cutinhead{The \SI{100}K expanded temperature scale \scaleone}
   TOP & 6354 &  4.01 & -2.23 &  2.26 &  5.80 &  4.59 &  3.05 &  3.41
\\ SGB & 5865 &  3.70 & -2.20 &  2.35 &  5.89 &  4.55 &  2.96 &  3.42
\\bRGB & 5486 &  3.36 & -2.15 &  1.41 &  5.88 &  4.58 &  3.02 &  3.41
\\ RGB & 5130 &  2.54 & -2.10 &  1.02 &  5.96 &  4.60 &  3.05 &  3.33
\\\cutinhead{The \SI{200}K expanded temperature scale \scaletwo}
   TOP & 6454 &  4.03 & -2.21 &  2.33 &  5.84 &  4.64 &  3.09 &  3.49
\\ SGB & 5925 &  3.72 & -2.18 &  2.41 &  5.92 &  4.59 &  2.99 &  3.48
\\bRGB & 5526 &  3.37 & -2.14 &  1.43 &  5.90 &  4.61 &  3.04 &  3.46
\\ RGB & 5130 &  2.54 & -2.10 &  1.02 &  5.96 &  4.60 &  3.05 &  3.33
\enddata
\tablecomments{
Uncertainties in elemental abundances on the differential scale are assumed to be \SI{0.05}{dex}. 
For iron, we use the statistical uncertainties of 0.04, 0.05, 0.04, and \SI{0.03}{dex} for the TOP, SGB, bRGB, and RGB, respectively, as determined in \citetalias{korn2007}.
For magnesium, we estimate the uncertainty to be \SI{0.065}{dex} for the TOP, SGB and bRGB groups, and \SI{0.05}{dex} for the RGB.
}
\tablenotetext{a}{\draftbf{In the spectroscopic analysis, \logg\ is corrected for the diffusion of helium by \SI{-0.05}{dex} for the TOP and SGB.}} 
\tablenotetext{b}{Magnesium has been corrected for pollution effects (see \secref{sec:pollution}).}
\end{deluxetable}

\subsection{Anticorrelations} \label{sec:pollution}
As noted in \citetalias{korn2007}, the bRGB and RGB groups show some scatter in the magnesium abundance. Magnesium is known to anticorrelate with aluminum in globular clusters \citep[see][]{gratton2004}, but no lines of the latter are detected in our spectra.
We resort to using sodium as a proxy via its well-studied O-Na anticorrelation \citep[e.g.][]{carretta2009, gratton2004}, and the related Li-Na anticorrelation which is known to be significant only for the most strongly polluted stars \citep{lind2009b}.

From a large sample of RGB stars, \citet{carretta2009} have determined that some $25 \pm \SI{13}{\%}$ of stars in \thecluster\ belong to the first generation. The independent determination by \citet{lind2011} confirms this result. 
They identify for their sample of turnoff stars mean abundances for the first and second generations as $\abundance(\elem{Na}) \approx 3.65$ and 4.05, respectively. For their combined sample of giants and diffusion-corrected dwarfs ($\SI{+0.20}{dex}$), the levels increase to $\abundance(\elem{Na}) = 3.84 \pm 0.10 $ and $4.23 \pm 0.13$, respectively.
As our sample size is insufficient for statistical analyses of the abundance trends, \draftbf{we rely on comparisons with the data and results of \citet{carretta2009} and \citet{lind2008,lind2009b,lind2011} for our population determinations}.
We perform the sodium abundance analysis on our original spectroscopic \Halpha\ temperature scale, as its good agreement with previous photometry makes it directly comparable with the cited studies, giving us comparable abundance scales.
This assumption is further supported by the agreement of stellar parameters for \draftbf{stars in common with the other studies}.

For our sample of six RGB stars, we find a strong anticorrelation between magnesium and sodium, identifying two out of our six stars as belonging to the first generation in this cluster with $\abundance(\elem{Na}) \approx 3.9$. Individual measurements of the very weak lithium line gives a relatively large scatter ($\sigma \approx \SI{0.10}{dex}$), seemingly anti-correlated to sodium abundances at rather low significance.

As indicated in \citetalias{korn2007}, our sample of bRGB stars shows a bimodal distribution of magnesium abundances ($\Delta \approx 0.17 \pm 0.05$). No anticorrelation is detected, as the very weak sodium line seems to exhibit only mild scatter ($\sigma \approx \SI{0.06}{dex}$), without correlation to the former. Applying different effective temperatures from various photometric calibrations does not alter this result. 
Lithium shows significant scatter in line with expectations from an evolutionary scatter within the group (the cooler stars have more strongly diluted surface layers, with lower lithium abundances). Absolute sodium abundances ($\abundance(\elem{Na}) \approx 4.2$) indicate that all five stars belong to the second generation.

Within the SGB and TOP groups, no significant scatter is detected for magnesium. Sodium abundances identify both SGB stars as borderline second generation ($\abundance(\elem{Na}) \approx 4.0$), while all five TOP stars clearly belong to the second generation ($\abundance(\elem{Na}) > 4.0$).

We account for pollution effects by attempting to restore the magnesium abundances of the first-generation stars. We have directly identified first-generation stars in the RGB group, while for the other groups we must extrapolate from comparisons with the aforementioned studies. We apply upward corrections by $0.08, 0.07, 0.11, \SI{0.11}{dex}$ to the group-averaged magnesium abundances for the RGB, bRGB, SGB, and TOP group, respectively. We propagate estimated uncertainties of the latter three corrections into those of the abundance determinations.

If better abundance determinations for e.g. sodium or nitrogen were available, one could in principle reject polluted stars in the abundance analyses of sensitive elements. Unfortunately, it seems none of our TOP and SGB stars belong to the first generation, and thus may well represent a slightly decreased abundance in lithium. The magnitude of this effect may be gleaned from the observed difference amongst dwarfs in the large sample of \citet{lind2009b}, $\le \SI{0.05}{dex}$.
Future studies could apply the identification method based on the nitrogen-sensitive photometric index $c_y$ when selecting their targets \citep[see][and references therein]{lind2011}.

\subsection{Abundance trends}


On our preferred temperature scale, \scaleshifted, $\Delta \Teff$ is not altered, thus affecting abundance trends only by second-order effects. For magnesium, the corrections for pollution weaken the abundance trend to $\Delta \abundance (\RGB-\TOP) = 0.18 \pm 0.08$, which is significant on the $2 \sigma$ level.
The trend in iron, meanwhile, is significant on the $3 \sigma$ level with $\Delta \abundance (\RGB-\TOP) = 0.16 \pm 0.05$.
Calcium and titanium both show mildly significant trends, $\Delta \abundance (\RGB-\SGB) = 0.12 \pm 0.07$. Lithium displays a significant upturn at the SGB, $\Delta \abundance (\SGB - \TOP) = 0.13 \pm 0.07$, in line with the independent results of \citet{lind2009b}.
The evolution of chromium is consistent with no variation.

\draftbf{Comparing to \citetalias{korn2007}, magnesium abundances have been most significantly altered by correcting for cluster-internal pollution. This decreases $\Delta \abundance(\RGB-\TOP)$ by \SI{0.03}{dex}, while increasing the uncertainty of this measure (from 0.07 to \SI{0.08}{dex}).
The altered temperature scale has not affected abundance trends in neither magnesium nor lithium. 
Iron abundances are now based upon ionic lines, which are not temperature sensitive.
The absolute abundances of iron and titanium are affected by the new, increased \logg\ values, with insignificant differential effects. The trend in calcium has strengthened by the combined differential effects from \Teff\ and \logg\ by \SI{0.04}{dex}.
}

\draftbf{Expanding the temperature scale, i.e. increasing parameter $\Delta \Teff (\TOP-\RGB)$, implies higher abundances for the TOP stars. This leads to higher inferred abundances in temperature sensitive elements, which decreases parameter $\Delta \abundance(\RGB-\TOP)$.
This tends to flatten the observed abundance trends in \citetalias{korn2007}, which all exhibited $\Delta \abundance(\RGB-\TOP) > 0$.
}

On the expanded temperature scale \scaleone, trends in magnesium and iron are both significant on the $2 \sigma$ level. 
Titanium and chromium exhibit similar but weakly significant trends of opposite sign. Calcium abundances are consistent with no variation. The upturn in lithium abundances at the SGB remains, $\Delta \abundance(\SGB-\TOP) = 0.10 \pm 0.07$.

On the hottest temperature scale, \scaletwo, abundance trends by construction flatten significantly.
The variation in iron however remains significant with $\Delta \abundance(\RGB-\TOP) = 0.11 \pm 0.05$.
Chromium too deviates strongly, due singly to the behavior on the RGB. We note that while most signatures flatten, $\Delta \abundance(\TOP-\SGB)$ of titanium actually increases somewhat on this temperature scale. This could be counteracted by arbitrarily increasing \logg\ values on the SGB, while inducing a complementary effect on iron.
Flattening the iron trend further would similarly require increasing \logg\ values at the TOP, with the drawback of equally strengthening the trend of titanium. Hence, this scenario results in nearly optimally flattened abundance variations.

\subsection{Comparison with diffusion models} \label{sec:diffusion}

\newcommand\figtitle[1]{\centerline{\LARGE #1} } 
\begin{figure*}[t]
\figtitle{The shifted \Teff\ scale \scaleshifted}
\includegraphics[angle=90,width=\linewidth]{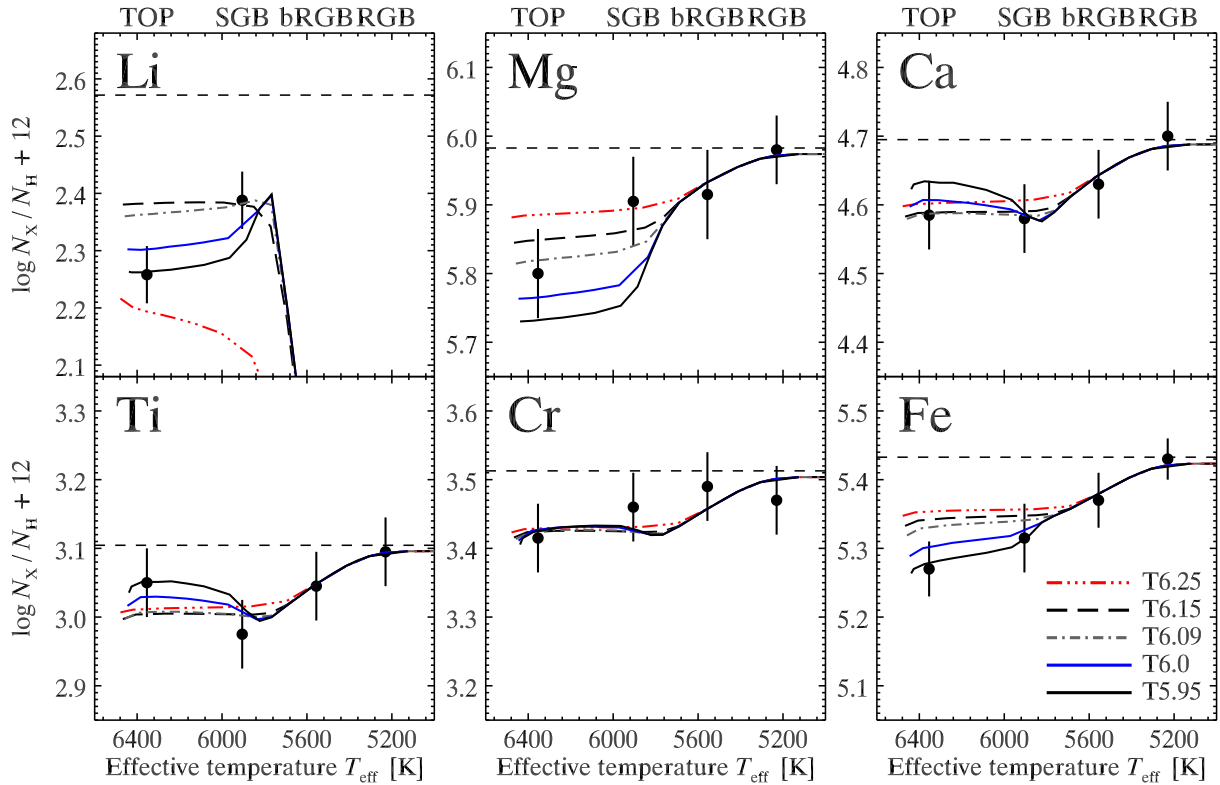}
\caption{Chemical abundances on the shifted temperature scale \scaleshifted, compared to predictions from \SI{12.5}{Gyr} isochrones. 
Dashed horizontal lines represent the initial composition.
The stellar models include the effects of atomic diffusion with a free parameter for the \tme, at five different values. We find that models in the range T5.95--6.15 all reproduce observations quite well. The optimal model is T6.0.
\coloredition\label{fig:trends-100s}}
\end{figure*}

\newcommand\imgscale{0.75}
\begin{figure*}[p]
\figtitle{The expanded \Teff\ scale \scaleone}
 \centerline{\includegraphics[angle=90,width=\imgscale\linewidth]{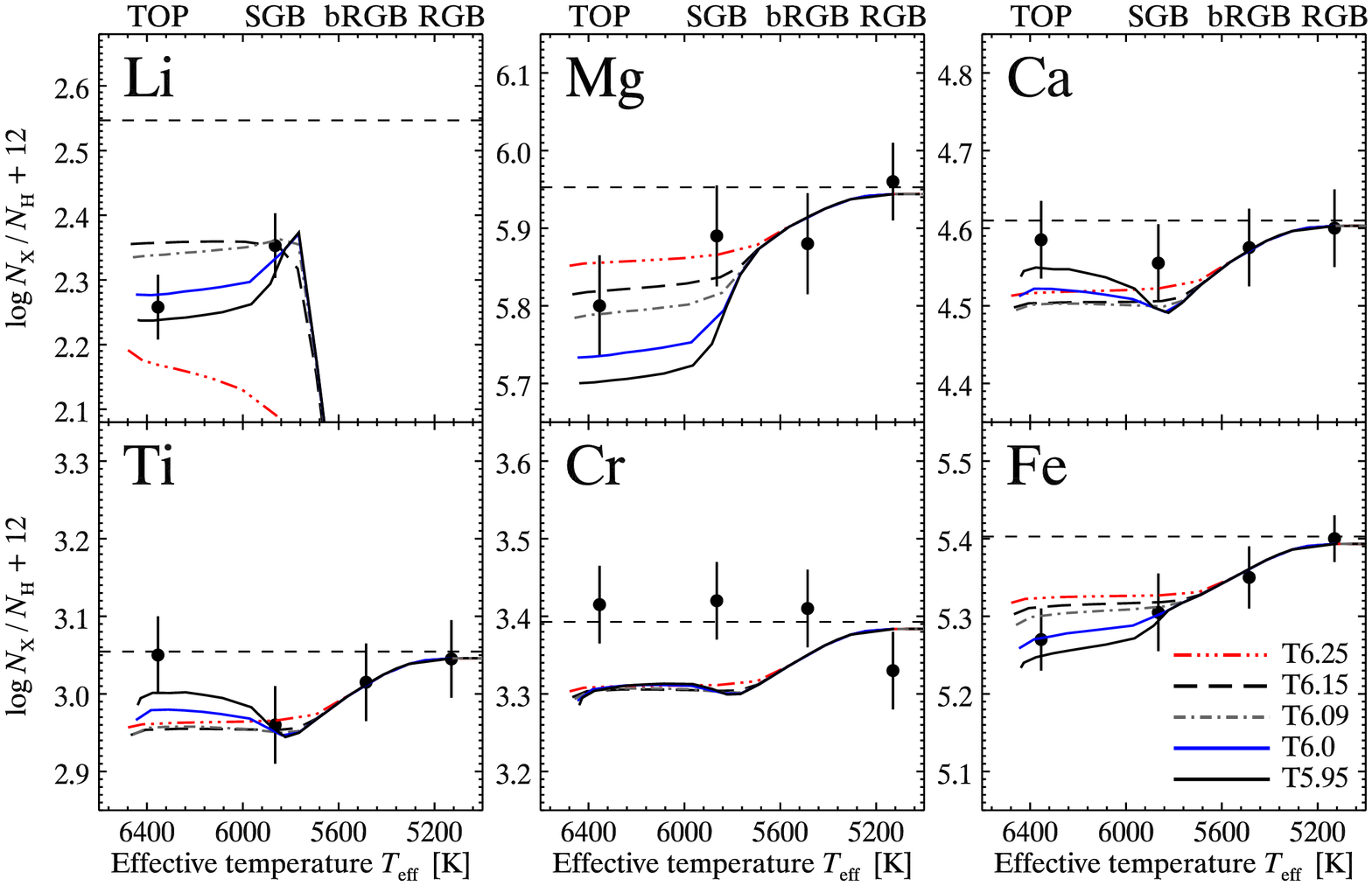}}
\caption{Chemical abundances on the expanded temperature scale \scaleone\ (see \tblref{tbl:params}), compared to predictions from \SI{12.5}{Gyr} isochrones.
Dashed horizontal lines represent the initial composition.
\label{fig:trends-100e}}
\end{figure*}
\begin{figure*}[p]
\figtitle{The hot expanded \Teff\ scale \scaletwo}
\centerline{\includegraphics[angle=90,width=\imgscale\linewidth]{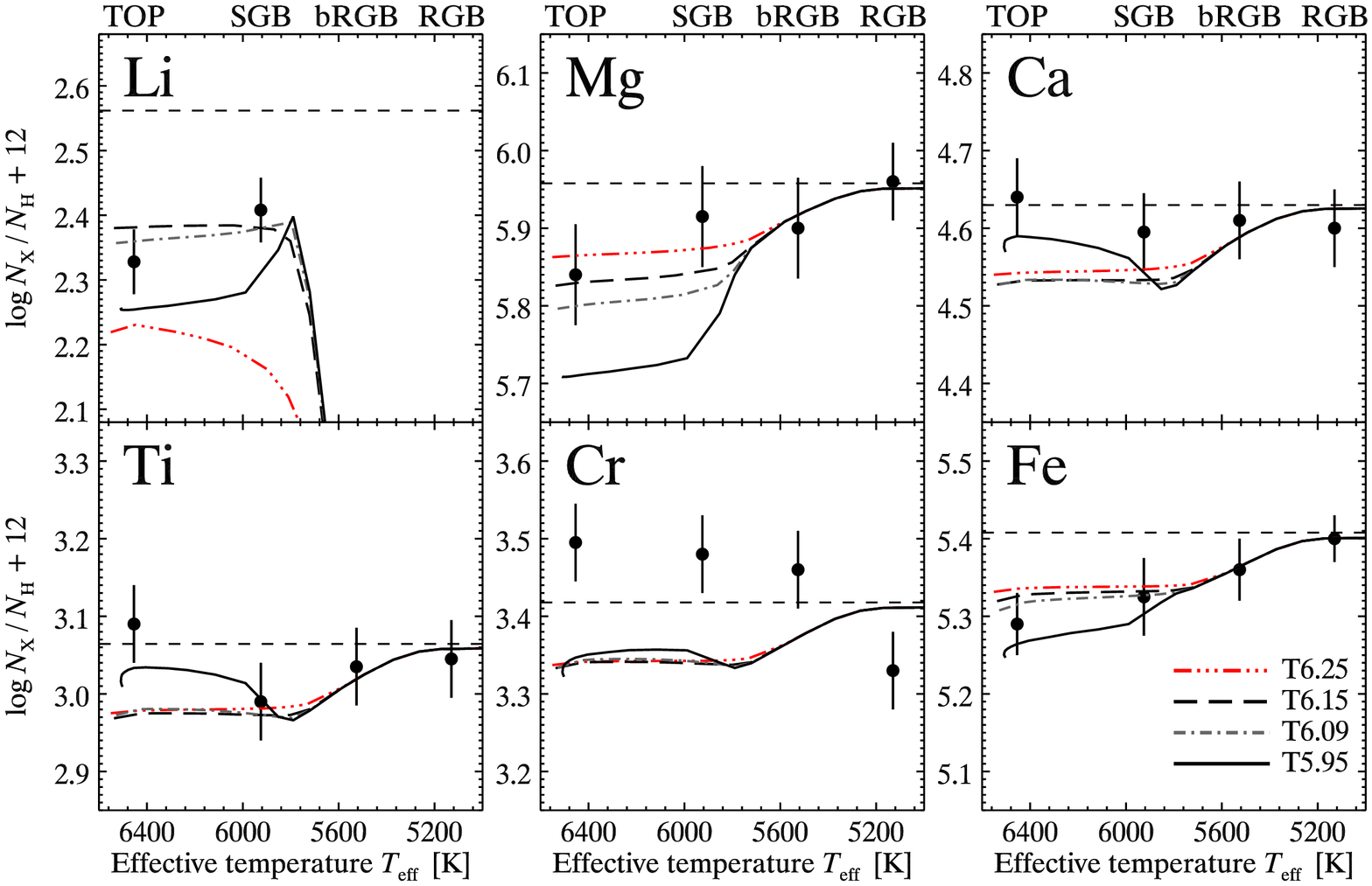}}
\caption{Chemical abundances on the hot expanded temperature scale \scaletwo\ (see \tblref{tbl:params}), compared to predictions from \SI{11.5}{Gyr} isochrones.
Dashed horizontal lines represent the initial composition.
\label{fig:trends-200e}}
\end{figure*}

In \figref{fig:trends-100s}, we compare observed abundance trends on the shifted temperature scale \scaleshifted\ to stellar-structure predictions, as introduced in \secref{sec:diffmethod}.
Results on the expanded temperature scales are presented in \figref{fig:trends-100e} and \figref{fig:trends-200e}.
The dashed horizontal line represents the initial composition of the models, adjusted to the observed abundance level.
The precise choice of age, 11.5 versus \SI{12.5}{Gyr}, does not significantly influence the strength of the predicted evolutionary trends. Differences are larger when comparing to \SI{13.5}{Gyr} models, in the sense that those predict stronger abundance trends for a given \tme.
\draftbf{The influence of age is most significant on the cluster CMD -- see \secref{sec:tscales}.
From the discussion there}, we compare \scaletwo\ to the \SI{11.5}{Gyr} isochrones, and the other temperature scales to the \SI{12.5}{Gyr} isochrones, where more models are available.

For the shifted temperature scale \scaleshifted, the models of least efficient turbulent mixing, T5.95--6.0, find excellent agreement with the observed trends in lithium, calcium, iron, chromium and titanium. For calcium, the predicted trend is somewhat stronger than observed, and for magnesium this is most apparent for the abundance at the SGB. 
Lithium, titanium and iron all exhibit trends favourable of weak efficiency. Taking into account the stronger upward adjustment required to match the TOP and SGB abundances of lithium, the disagreement with the more uncertain RGB stars lessens when the T6.0 models are chosen. Hence, as in \citetalias{korn2007}, this remains the optimal model, which gives the best overall agreement with observations.

Models with moderately strong efficiency, T6.09 and T6.15, find mild disagreement with the predicted upturn in lithium abundances at the SGB, where instead a flat behavior is predicted. Similarly, the evolutionary trend of iron becomes too weak. At less than $2 \sigma$ disagreement for all elements, these models can, however, not be excluded.

\draftbf{Models with very strong} \tme, T6.25, disagree with the evolutionary trend of iron on the $2 \sigma$ level. 
Even worse, they disagree on the $3 \sigma$ level with the evolutionary trend of lithium, where a strong downturn is predicted at the SGB. 
\draftbf{This is because, with turbulent mixing this strong}, gravitationally settled material is brought to sufficiently hot layers that lithium is destroyed. As the surface convection zone expands inward, it mixes with depleted material, thus diluting the surface abundance further. Notably, this causes additional \SI{0.15}{dex} of lithium destruction, which naturally carries into the evolution along the RGB.

Hence, models with \tmes\ in the range T5.95--6.15 agree well with observations, with a preference for the weaker efficiencies. A strong \tme, T6.25, can be excluded for our sample of stars at this temperature scale from its predictions for both iron and lithium.

On the expanded temperature scale \scaleone, the above results hold, with a possible preference for slightly stronger \tme. The sole exception is chromium, where the TOP--bRGB range exhibits no variation, but $\Delta \abundance(\TOP-\RGB)$ disagrees on the $2 \sigma$ level, with similar magnitude but opposite sign to predictions.
\draftbf{This result may however be significantly altered by \meanthreed\ NLTE modeling (see \secref{sec:nlte}).}

On the strongly expanded temperature scale \scaletwo, trends by design flatten further, while not significantly altering the above conclusions. A weak efficiency of turbulent mixing is still compatible with all observed trends, except for magnesium which clearly prefers somewhat higher efficiency, and chromium, discussed previously.

\section{Discussion} \label{sec:discussion}
We have updated the analysis of \citetalias{korn2007} by deviating from its internally consistent spectroscopic stellar parameters, and instead moving toward results inspired by the photometric calibrations by \citet{casagrande2010}. 
Further, we have corrected the abundances of magnesium \draftbf{on the grounds of two stellar generations in the globular cluster, as identified by \citet{carretta2009} and \citet{lind2009b,lind2011}}. Only two of our stars may be easily identified as belonging to the first, pristine, generation, while the others seem to belong to a second generation with atmospheres to some degree polluted by more massive stars belonging to the first.

We have constructed three new temperature scales. 
Two of these find good agreement with our spectroscopic and photometric indicators: the shifted \scaleshifted\ and the expanded \scaleone\ temperature scales. 
Both are compatible with the WDCS cluster age of $\SI{12.0 \pm 0.5}{Gyr}$ \citep{hansen2007,kowalski2007}.
They exhibit evolutionary abundance trends, significant on the $2 \sigma$ level. On the shifted \scaleshifted\ temperature scale, the abundance trend in iron is significant at $3 \sigma$. These results are compatible with predictions from stellar-structure models considering atomic diffusion moderated by turbulent mixing.
\draftbf{Allowed models use turbulent mixing of weak to moderate efficiency. The optimal model is found to be the same as identified in \citetalias{korn2007}.
The high-efficiency models preferred elsewhere \citep{gonzalez2009,melendez2010} 
produce additional destruction of lithium, which leads to strong disagreement between our inferred abundances at the TOP and SGB.
}

The third temperature scale, \scaletwo, was constructed under the criterion of optimally flattening abundance trends. This is achieved, with the exceptions of iron and chromium at the $2 \sigma$ level (we dismiss the latter due to the large 3D corrections known to exist -- see \secref{sec:nlte}). 
From the point of view of atomic diffusion however, we recognize systematic albeit weak abundance trends in good qualitative accordance with predictions for all elements but chromium.
Additionally, this temperature scale violates our spectroscopic effective temperatures from \Halpha\ analyses, as well as all photometric calibrations considered.
Finally, our magnesium abundances were corrected for pollution signatures deduced from sodium abundances. These were identified using the results of \citet{lind2011}, where corrections for diffusion were applied. Neglecting this effect and instead calibrating on their large sample of dwarf-star sodium abundances results in a strengthened magnesium abundance trend significant at the $2 \sigma$ level. In this sense, it is not at all clear that there exists a temperature scale which does away with the need for atomic diffusion altogether. 

Finally, we acknowledge the differential analysis of TOP and cool RGB stars in \thecluster\ by \citet{koch2011}.
As our RGB stars have already undergone sufficient dilution of surface layers to restore all abundances but lithium to their original levels -- signified by dashed lines in \figref{fig:trends-100s} -- one would not expect stronger effects to appear for their cooler stars. 
From their Fig.~4, with $\Delta \equiv \Delta({\rm RGB}-\TOP)$, their absolute abundance variations in magnesium ($\Delta \approx \SI{0.16}{dex}$) and iron ($\Delta \approx \SI{0.14}{dex}$, based on \ion{Fe}2) agree very well with ours, as well as model predictions. Their results for titanium ($\Delta \approx \SI{0.28}{dex}$, based on \ion{Ti}2) and calcium ($\Delta \approx \SI{-0.11}{dex}$) deviate in alternate directions by $\SI{0.2}{dex}$.

\subsection{The primordial lithium abundance}

\begin{figure}[t]
 \includegraphics[angle=90,width=\linewidth]{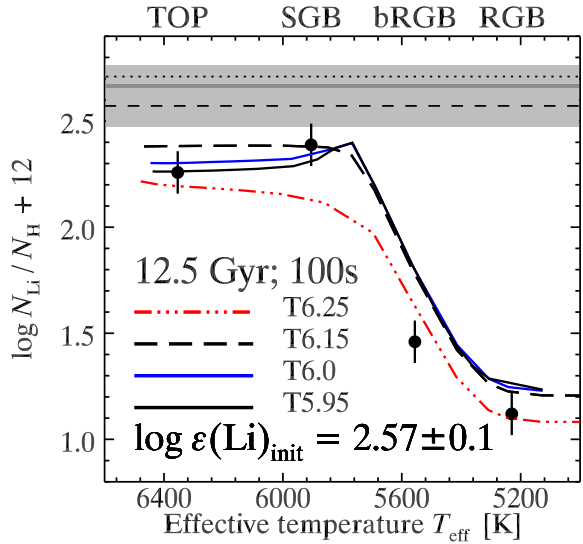}
 \caption{Observed abundance trends in lithium, shown for the shifted temperature scale \scaleshifted.
Lines represent the predicted surface abundance evolution at four different efficiencies of turbulent mixing. 
The horizontal dashed line represents the original lithium abundance of all four models, the dotted line the predicted primordial abundance. The shaded areas represent the respective uncertainties, and their overlap.
\coloredition\label{fig:lithium}}
\end{figure}

In \figref{fig:lithium}, we show that 
the initial lithium abundance of the cluster as inferred from our optimal model (T6.0), $\abundance(\elem{Li})_{\rm init} = 2.57 \pm 0.1$, is compatible with the predicted primordial abundance based on \wmap\ data \citep[$\abundance(\elem{Li})_{\rm BBN} = 2.71 \pm 0.06$]{cyburt2010}. Adopting the mutual error bars, the values differ by $\SI{0.14 \pm 0.11}{dex}$, or $1.2 \sigma$. 
If instead the T5.95 model were adopted, the initial abundance would further increase by \SI{0.04}{dex} to $\abundance(\elem{Li})_{\rm init} = 2.61$. Modeling in \meanthreed\ NLTE may further increase the abundance slightly, as would the assumption of some cluster-internal lithium destruction. 
\draftbf{Adopting a cluster age of 12.0 or \SI{11.5}{Gyr} decreases the predicted surface depletion slightly, which reduces the inferred initial abundance by at most \SI{0.01}{dex}.}

\draftbf{The model of high-efficiency model (T6.25)} is not compatible with the lithium abundances on the TOP and SGB. The deep mixing causes additional destruction of lithium by \SI{0.15}{dex}. When calibrating solely on the RGB, this would result in the inferred initial abundance $\abundance(\elem{Li}) = 2.60$. Note however that lithium abundances on the RGB are uncertain from the modeling perspective as regards both line formation theory (see \secref{sec:nlte}) and stellar evolution models (as indicated by a possible offset in \Teff\ or insufficient depletion of lithium along the subgiant branch, as seen in \figref{fig:lithium}, and likewise in \citet{lind2009b}). 
These problems must be taken into account when considering the analysis of \citet{mucciarelli2012}. Indeed, these stars are less sensitive to the precise efficiency of atomic diffusion. But additional destruction of lithium occurs in models with strong turbulent mixing, an effect also affecting RGB stars, leading to underestimated initial abundances.
We note in passing that their mean lithium abundance on the RGB for this cluster, $\abundance(\elem{Li}) = 1.09 \pm 0.1$, agrees with ours on the temperature scale \scaleshifted, $\abundance(\elem{Li}) = 1.12 \pm 0.1$.

\section{Conclusions}

\draftbf{We find agreement between the stellar parameters derived on two temperature scales compatible with the \citet{casagrande2010} calibration, and the stellar models with atomic diffusion of \citet{richard2005} assuming an age for NGC~6397 of $12.0 \pm \SI{0.5}{Gyr}$, as determined by \citet{kowalski2007}.
The spectroscopically determined abundances of six chemical elements on these temperature scales exhibit significant variations, in agreement with predictions from the stellar models if a weak efficiency of turbulent mixing is assumed. 
These results are rather robust as regards the choice of temperature scale.
We cannot reconcile the detected abundance variations with predictions from models using the strong \tme\ identified as optimal for Spite plateau stars in the field \citep{melendez2010}.
}

\draftbf{On the preferred temperature scale, the optimal model indicates an initial lithium abundance of $\abundance(\elem{Li})_{\rm init} = 2.57 \pm 0.1$. 
This is compatible with the predicted primordial abundance based on \wmap\ data \citep[$\abundance(\elem{Li})_{\rm BBN} = 2.71 \pm 0.06$]{cyburt2010}.
}

We note the WMAP-independent primordial lithium abundance of \citet{steigman2010}, $\abundance(\elem{Li}) = 2.65 \pm 0.06$, in excellent agreement with our results.
This value is computed from the baryon density compatible with deuterium abundances in high-redshift QSO absorption line systems.
Additionally, one should expect some reduction from the primordial abundance due to mixing through population III stars \citep[a mild version of the scenario presented by][]{piau2006}.
Finally, the first and second generation stars in this cluster may exhibit slightly different lithium abundances. Previous studies \citep{lind2009b} identify lithium abundances higher in first than second generation dwarfs by $\le \SI{0.05}{dex}$. \draftbf{We note that a differential analysis comparing our TOP stars to the field star \hdref\ implies equal lithium abundances.}
Future studies could focus on analyses of first generation stars, to remove this minor possible bias when constraining the true initial lithium abundances of cluster stars.

In view of these uncertainties on either side, one may call into question the significance of the remaining cosmological lithium discrepancy. We can, however, be more certain about the reality of atomic diffusion in stars on the traditional Spite plateau, as shown by theoretical and observational investigations alike.
\draftbf{What is still missing is an understanding of the physical processes that give rise to the mixing required to moderate atomic diffusion.
We call on stellar-structure experts to intensify investigations into possible (hydrodynamic) effects.}

\acknowledgments
We wish to thank E. Carretta for providing us with supplemental UVES spectra,
M. Bergemann for chromium NLTE corrections and informative discussions,
and P. S. Barklem for his \Halpha\ analyses.
We would also like to thank B. Edvardsson and B. Gustafsson for valuable discussions.
O. R. thanks the R\'eseau Qu\'eb\'ecois de Calcul de Haute Performance (RQCHP) and the Centre de Comp\'etences en calcul haute performance de la r\'egion Languedoc-Roussillon (HPC@LR) for providing the computational resources required for this work.
A. J. K. acknowledges funding through the Swedish National Space Board and the ESF EuroGENESIS program.
This research has made use of the VizieR catalogue access tool, CDS, Strasbourg, France. 

{\small

}

\end{document}